\documentclass[sigconf,authorversion,nonacm]{acmart}
\usepackage{multirow}

\AtBeginDocument{%
  \providecommand\BibTeX{{%
    \normalfont B\kern-0.5em{\scshape i\kern-0.25em b}\kern-0.8em\TeX}}}
\begin{document}
\title[Breaking the Curse of Quality Saturation with User-Centric Ranking]{Breaking the Curse of Quality Saturation with User-Centric~Ranking}
\titlenote{Accepted to publish at the 29th ACM SIGKDD Conference on Knowledge Discovery and Data Mining (\emph{KDD' 2023}).}

\author{Zhuokai Zhao}
\email{zhuokai@uchicago.edu}
\affiliation{%
  \institution{University of Chicago}
  \city{Chicago}
  \state{IL}
 \country{USA}
}
\authornote{Work done during an internship at Meta.}

\author{Yang Yang}
\email{yzyang@meta.com}
\affiliation{%
  \institution{Meta AI}
  \city{Menlo Park}
  \state{CA}
 \country{USA}
}

\author{Wenyu Wang}
\email{owenwang@meta.com}
\affiliation{%
  \institution{Meta AI}
  \city{Menlo Park}
  \state{CA}
 \country{USA}
}

\author{Chihuang Liu}
\email{chihuang@meta.com}
\affiliation{%
  \institution{Meta AI}
  \city{Menlo Park}
  \state{CA}
 \country{USA}
}

\author{Yu Shi}
\email{yushi2@meta.com}
\affiliation{%
  \institution{Meta AI}
  \city{Menlo Park}
  \state{CA}
 \country{USA}
}

\author{Wenjie Hu}
\email{wenjiehu@meta.com}
\affiliation{%
  \institution{Meta AI}
  \city{Menlo Park}
  \state{CA}
 \country{USA}
}

\author{Haotian Zhang}
\email{htzhang@meta.com}
\affiliation{%
  \institution{Meta AI}
  \city{Menlo Park}
  \state{CA}
 \country{USA}
}

\author{Shuang Yang}
\email{shuangyang@meta.com}
\affiliation{%
  \institution{Meta AI}
  \city{Menlo Park}
  \state{CA}
 \country{USA}
}
\authornote{Lead author; part of the work was done at NewsBreak before joining Meta.}

\renewcommand{\shortauthors}{Zhao et al.}

\begin{abstract}
  A key puzzle in search, ads, and recommendation is that the ranking model can only utilize a small portion of the vastly available user interaction data. As a result, increasing data volume, model size, or computation FLOPs will quickly suffer from diminishing returns. We examined this problem and found that one of the root causes may lie in the so-called ``item-centric'' formulation, which has an unbounded vocabulary and thus uncontrolled model complexity. To mitigate quality saturation, we introduce an alternative formulation named ``user-centric ranking'', which is based on a transposed view of the dyadic user-item interaction data. We show that this formulation has a promising scaling property, enabling us to train better-converged models on substantially larger data sets.
\end{abstract}

\ccsdesc[500]{Information systems~Recommender systems; Personalization}

\keywords{Ranking; Recommendation Systems; Collaborative Filtering; Deep Learning; User-centric Ranking }



\maketitle
\section{Introduction}
Scaling has been one of the main themes in deep learning and the key driving force behind many eye-opening breakthroughs in the past decade, especially in computer vision (CV)~\cite{dos2021ViT,feich2022MAE,wang2022git}, natural language processing (NLP)~\cite{devlin2018bert,brown2020GPT,chowdhery2022palm}, and multi-modality modeling~\cite{ramesh2021DALLE,yu2022coca,radford2021CLIP}. In these areas, scaled-up big models were able to improve the corresponding quality metrics by orders of magnitude compared to the state-of-the-art of their previous generations. For example, on ImageNet~\cite{deng2009imagenet}, the ViT~\cite{dos2021ViT} model reduced the image classification error rate, compared to the first super-human model ResNet-152~\cite{he2016deep}, by more than half~\cite{dos2021ViT}. This scaling success, however, has not yet happened in ranking (e.g., search, ads, recommendation systems). This seems both surprising and mysterious given that ranking is by far the most incentivized application in the AI industry.


In a typical scaling scenario, one important condition is that the model should have the capability to utilize more data, so that increasing data volume and computing will continue to improve model quality. When it comes to ranking, we notice that even with an abundant or even infinite amount of data (i.e., massive user engagement activities constantly accumulating in systems like Google ads, 
Facebook news feed, YouTube video recommendation, etc.), the ranking models typically can only utilize a small portion (i.e., a few days to a few weeks of logged data). Increasing training data volume, model size, or computation FLOPs can only lead to very little quality improvement. This is known as the ``quality saturation'' problem. 

To be fair, the quality of every machine learning model will eventually saturate, sooner or later. What makes it unique in ranking is that the quality saturation happens too soon. Considering the important role that ranking models play and their business impact, a reasonable expectation is that a ranking model should be able to utilize at least a few months of training data.

We examined this problem and found that one of the root causes may lie in the formulation. With an analogy to NLP, the current ranking formulation predicts dyadic responses (e.g., ads click-through) by casting `items' as `tokens' and `users' as `documents', a paradigm called ``item-centric ranking''. This is actually an ill-posed formulation because the model size or the number of parameters to learn will grow linearly as data volume increases. 
As a remedy, we introduce an alternative formulation called ``user-centric ranking'' based on a transposed view, which casts `users' as `tokens' and `items' as `documents' instead. We show that this formulation has a number of advantages and shows less sign of quality saturation when trained on substantially larger data sets.  

The proposed methods have been tested in a variety of our production systems with significant metric wins, including search, ads, and recommendation. These systems are quite diverse in nature (e.g, different interaction interfaces, items of very different types) and can be regarded as representative of many ranking systems in the industry, yet our findings are quite consistent. Our reported experiment results are primarily based on one production surface, which has 6 different tasks (including both positive and negative engagements, and both immediate and deferred reward feedback), and the comparison and trend are consistent across all these tasks. In addition to offline results, we also report online live experiment results. Furthermore, to improve the reproducibility of our findings, we also include results on a public data set and plan to open-source our implementation code for public access.


\begin{figure*}[htbp]
\centering
\begin{tabular}{ccc}
\includegraphics[height=0.42\linewidth]{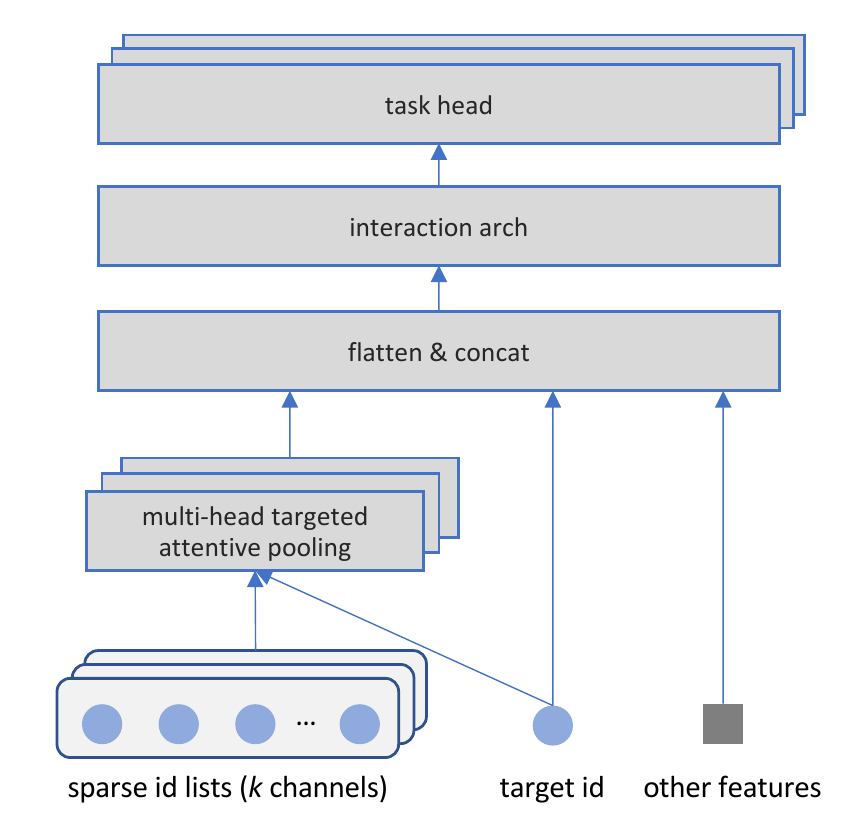} & & \includegraphics[height=0.42\linewidth]{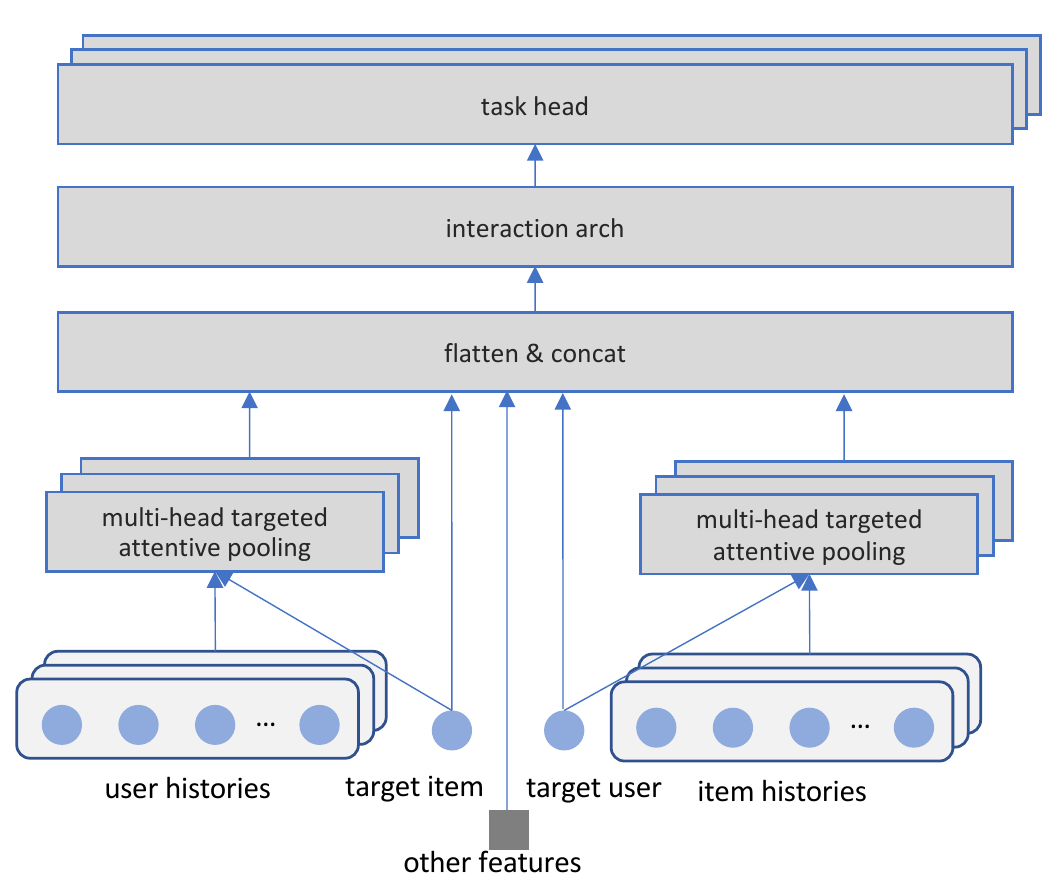}\\
~~~~~~~~(a)  & & ~~~~~~~(b)
\end{tabular}
\vspace{10pt}
\caption{(a) An example of one-tower ranking model; (b) A hybrid ranking model containing both a user-centric and an item-centric sub-architecture.}
\label{FigureICR}
\end{figure*}

\section{Related Work}
The past decade has witnessed tremendous successes achieved by deep learning models that are growing in scale exponentially over time. In computer vision, big model architectures have been widely used for image classification and object detection tasks. The neural architectures have evolved from Convolutional Neural Networks (CNNs) with a handful of layers~\cite{NIPS2012_c399862d}, to ResNet who has more than 100 layers and 100 million parameters~\cite{he2016deep}, to recent gigantic Transformer-based models that contain hundreds of billions of parameters~\cite{dos2021ViT}. The trend is even more prominent in NLP, especially in the few years of post-BERT era~\cite{vaswani2017attention, devlin2018bert}. A surge of state-of-the-art models are emerging with ever growing sizes, complexities, and new levels of capabilities, e.g, GPT-3 and GLaM \cite{du2022glam} are among the largest language models to date and have demonstrated impressive performance in various NLP tasks~\cite{brown2020GPT,chowdhery2022palm}.


It is a bit surprising that, unlike the other areas, scaling has not gained much success in ranking, even though it is the biggest industry for AI and there is no shortage of training data~\cite{Dacrema}. Ranking models used to be dominated by the ``two-tower'' architectures, where the user-side and the item-side were modeled independently with separate architectures in the early stage known as the two towers; and fusion or interaction between the two sides happens at a relative late stage~\cite{cheng2016wide,huang2013dssm,he2017neural}. Recently, ``single-tower'' architectures based on Transformer emerged and quickly became the new state of the art~\cite{vaswani2017attention,zhou2018deep}. However, compared to other areas, these models are notably simpler, for example, they are using only a single (or a few, if Transformer is also used in interaction sub-arch) layer of Transformer block, and even though these models could be big in size (e.g, 1 trillion parameters), the majority of the parameters are sparse-id based embeddings, only a tiny fraction of which are active for each prediction.

The current common practice in ranking is to model each user based on the sequence of historically interacted items. The representation of user interests can be learned from historical behaviors, and the likelihood of a potential engagement is assessed based on the affinity of the target item with respect to historical interactions. These models provide an item-centric perspective to utilize the dyadic user-item interaction data; we call it item-centric because learnable embeddings are allocated for items but not users. We show that this formulation could be the cause of quality saturation. The proposed user-centric ranking is the first to provide an alternative formulation based on a transposed view of the dyadic interactions. We show that it can help to alleviate quality saturation in ranking. 
We want to note that our contribution is to introduce this new formulation, not a specific neural architecture. These two are orthogonal, in fact, any SoTA item-centric ranking model can be converted to its user-centric counterpart using the new formulation.

It is important to capture the complex relationships between users and items to improve ranking accuracy in ranking systems. Using user information corresponding to a target item is a natural choice. One example is graph-based recommendation models~\cite{wang2019neural, chen2020revisiting}, which represents users and items as nodes in a bipartite graph. The graph model learns to generate user and item embeddings for recommendation through the process of embedding, propagation, and prediction. Our approach of user-centric ranking models user-item interaction in a different way and targets for replacing or complementing the current item-centric ranking models that suffer from quality saturation. There are other attempts to alleviate the changing inventory problem, such as meta learning approaches~\cite{carvalho2008meta, wang2022meta}. The goal of meta learning for ranking is to improve robustness and/or fairness of ranking models caused by unintended data biases. In contrast, we aim to adddress the quality saturation problem caused by inventory dynamics.

\section{Ranking Formulations}
In ranking, we are concerned with modeling \textit{dyadic responses}.
Given a set of users $\mathcal{U}$ and a set of items $\mathcal{I}$, the goal is to predict $y_t(u,i)$ for any given user $u\in \mathcal{U}$ and item $i\in \mathcal{I}$ at time $t$. In different contexts, $y$ can have different semantic meanings, e.g., click-through of an ad, conversion of a transaction, following an account, or finishing watching a video. Ranking models are trained on historical interaction data in the format of $\mathcal{D} = \{(u,i,t, y)\}$, which can be thought of as a bipartite graph between $\mathcal{U}$ and $\mathcal{I}$.

An interesting note is that ranking bears a lot of similarities with NLP, because NLP data can be thought of as dyadic interactions between `documents' and `tokens'. 
In fact, a lot of ranking techniques are inspired by progresses in NLP~\cite{sun2019bert4rec,hidasi2015session,zhou2018deep}. 


\subsection{Item-Centric Ranking}\label{subsec:icr}
Figure~\ref{FigureICR} shows one example of single-tower item-centric architectures. The key idea, with an analogy to NLP, is to think of items as tokens and users as documents, i.e., each user is modeled by a list of items that they engaged with, in chronological order according to the time of engagements. When multiple types of engagements are involved (e.g., in video recommendations, engagements could include clicks, video completion, likes, follow-author, etc.), they can be organized into multiple channels, one for each engagement type. 

For each channel, items in the engagement history are first mapped to their embeddings, positions are encoded based on relative time-stamps, and multi-head attentions are applied on top. The aggregation output is then concatenated with all other features, on top of which an interaction sub-architecture (e.g., Deep \& Cross Network (DCN)~\cite{wang2021dcn} or self-attention~\cite{vaswani2017attention}) is employed to encode higher-order nonlinear interactions among different feature groups. And finally, a number of task heads (e.g., one MLP for each engagement prediction task) provide the output probabilities. Because of the daunting scale in ranking, these ranking architectures are highly-simplified versions compared to what are commonly used in NLP, noticeably: 1) only one layer of attention is typically used; 2) instead of full-sized self-attention, the aggregation is based on the so-called ``targeted attentive pooling", i.e., when predicting $y_t(u,i)$, the engagement history of user $u$ is aggregated by attending only w.r.t. the target item $i$ (i.e., the embedding of item $i$ is used as query in the attention function). The latter is similar to document/paragraph representation in NLP, where the aggregation is by attending to the special symbol `CLS'.

This formulation is called ``Item-Centric Ranking'' (ICR) to reflect that items are allocated free-parameter embeddings to be learned in training whereas user embeddings are derived by aggregating item embeddings. 

\subsection{User-Centric Ranking}\label{subsec:ucr}

Why do ranking models saturate so fast? Why doesn't this happen to NLP models given that they bear lots of similarities? When we carefully compare these two settings, we notice an important difference. In NLP, the vocabulary size (i.e., total number of tokens) is often fixed; given a neural architecture, the number of parameters is constant when we increase the training data. This is, however, not the case in ranking when item-centric formulation is used. 

In particular, especially in the so-called ``creator economy'', where the inventory of items are highly dynamic: new items are being created constantly (e.g., tens of millions of posts/videos are created on Facebook/Instagram every day) and items are time-sensitive and ephemeral (e.g., each post/video has a short life-span ranging from a few days to a few weeks). In this setting, because the item inventory grows linearly over time $|\mathcal{I}|=O(t)$, for any given neural architecture, the number of model parameters will grow unboundedly in $O(t)$ (due to the use of per-item embeddings). As a result, when we increase the training data (e.g., to use more days of logged interactions), because of the linear growth in model size, the per-parameter data density will not grow, and hence using more data will not make the model converge better (e.g., lower the variance). In fact, this is a setting that we rarely see elsewhere. 

Based on this observation, we propose an alternative formulation called ``User-Centric Ranking'' (UCR), which is based on a transposed view of the user-item interactions. Using the NLP analogy again, UCR casts `users' as `tokens' and `items' as `documents'; free-parameter embeddings are learned for users, and item embeddings are derived by aggregation. For mature ranking systems in double-sided markets, it is typical to see an increase in inventory, while the user set $\mathcal{U}$ remains relatively consistent; thus, the model size (i.e., the number of parameters) of these ranking systems will stay stable as we increase training data. Our expectation is that with this formulation, when we scale up training data the consistent growth of per-parameter data density should translate to better model convergence.

In a typical setting where user set is capped while both the inventory size and the training data set size  grow linearly over time, it can be shown 
the asymptotic error rate (i.e, the expected distance between the optimal value of model parameter $\theta^*$ and its actual value $\hat{\theta}$) for each of the formulations is as follows \cite{nguyen2018sgd}:
\begin{itemize}
    \item Item-centric ranking: $\mathbb{E}[||\theta^* - \hat{\theta}_t ||^2] = \text{Const}$ 
    \item User-centric ranking: $\mathbb{E}[||\theta^* - \hat{\theta}_t||^2] = O(\frac{1}{t})$ 
\end{itemize}
As training data grow, asymptotically UCR converges at a sublinear rate (at most), while ICR cannot be improved further, which explains the quality saturation we have observed.

From an intuitive perspective, UCR could be advantageous over ICR. In ICR, because items are ephemeral, so are their embeddings (i.e., an item embedding will soon become irrelevant and useless as that item exits the system). In UCR, we are continuously accumulating and improving our knowledge about every user by refining its embedding over time as long as that user keeps on interacting with the system. 

Any SoTA item-centric ranking model can be converted to its user-centric counterpart using the new formulation. Note that the example architecture in Figure 1(a) applies to both item-centric and user-centric. The key difference is whether users or items are used as keys for embedding look-ups (i.e, the `sparse-id' and `target-id' in the figure).

\subsection{Hybrid Models}\label{subsec:hybrid}
It is also possible and actually straightforward to have a hybrid formulation, i.e., to implement models that include both a user-centric and an item-centric attentive pooling components. Figure~\ref{FigureICR}(b) shows how the example architecture in Figure~\ref{FigureICR}(a) looks like in the hybrid formulation. Such hybrid models will have similar ``parameter explosion'' problem as item-centric models. We will compare all these different model formulations in our experiments.

\section{Implementation}\label{subsec:implementation}
\subsection{Item-Centric Ranking}\label{subsubsection:icr}
Item-centric id-lists represent the engagement history of each user. Although the number of items that one user can interact within one day is hardly over a few hundreds, the list of distinctive items and their embeddings gets accumulated very quickly over time, especially considering that the same item is rarely recommended to the same user again. A sampling strategy is needed in order for each engagement list to not exceed certain length. In our implementation, we limit the length to $1024$ at max, by only including the most recent engagements. In our experiment, this method is referred to as ``\texttt{IC-Sampling}''.

\subsection{User-Centric Ranking}\label{subsubsection:ucr}
One of the challenges for implementing UCR is to handle the distribution skewness. In an item-centric setting, the number of items one user can interact with tends to be evenly distributed (e.g., daily engagements range from a few to a few hundred), whereas in the new setting, the distribution is more irregular, e.g., some items can attract millions of users to engage with while others can get only a few.
This means that for some items it is no longer feasible to fit the entire list of engaged users in memory during training/inference. We explore three different approaches:



\begin{itemize}
    \item \textbf{Sampling}. In this implementation, we simply down-sample the list of engaged users of an item to a fixed-size sub-list uniformly using reservoir sampling.  Note that in practice, if we sample for each item only once, instead of resampling for each user-item interaction, this will introduce an artificial bias. This method is referred to as ``\texttt{UC-Sampling}."
    
    \item \textbf{Aggregation}. Another approach is to summarize a long sequence of engaged users to a shorter list, e.g., by clustering the users and using cluster-id in replacement of user-id. In our implementation, the clusters are obtained by applying the 
    Louvain algorithm \cite{blondel2008fast} to the user-item interaction graph. Our in-house implementation provides the functionality to incrementally update the clustering structure over time with constraints on cluster size and re-mapping ratio. This method is referred to as ``\texttt{UC-Clustering}".
    
    \item \textbf{Retrieval}. Alternatively, we can pre-index the engagement history and use retrieval (e.g., max inner-product search) to identify the subset of most relevant users (w.r.t. the target user), on which attentive pooling is then applied. Since attention is of quadratic complexity, the overhead of retrieval can be compensated by the speedup due to a shorter and more selective attention window. A sparsified attention distribution also means an improved signal-to-noise ratio (i.e., long-tail less relevant candidates are pruned and excluded from the attentive aggregation) and can further improve model quality. We leave this method for future investigation.	
\end{itemize}

Note that this problem is only a concern for a very small subset of the most popular items, for which most ranking models already have good prediction accuracy. For the vast majority of items in our case, the engagement users are below the 1024 length limit.

\subsection{Parameter Hashing}\label{subsubsec:parameter_hashing}
Another technical challenge is memory management when working with large-scale ID spaces such as user-ids $\mathcal{U}$ and item-ids $\mathcal{I}$. Considering that we are learning embedding vectors, one for each distinctive ID, the extremely large cardinalities (i.e., in the order of billions) of these ID spaces imply that the memory requirement as well as the index to map IDs to their address can be quite a challenge. Especially for item-centric ranking, the number of item IDs can grow unboundedly to infinity. 

One common approach to address this problem is to implement feature hashing, i.e., to maintain a constant hash space for these IDs and allocate one embedding vector for each distinctive ``hashed ID". This is of course not ideal. The existence of hash collisions means that we are forcing certain random IDs to share the same embedding vectors. This is not necessarily a bad thing when the collision rate is at a reasonable level, because feature hashing provides a type of regularization effect to the embedding parameters similar to dropout. However, for unbounded ID spaces such as  $\mathcal{I}$ in user-centric ranking, the collision rate is expected to grow linearly over time (i.e, $O(t)$), and can be arbitrarily large and no longer negligible. In contrast, in user-centric ranking, the ID space $\mathcal{U}$ is bounded and hence collision rate is under control.

\subsection{Aggregation Operators}\label{subsubsec:operators}
We implement two aggregation operators, sum-pooling and targeted attentive pooling. The former aggregates the list of associated IDs by the sum or mean of their corresponding embeddings. Sum-pooling is computationally inexpensive and easy to implement. However, it has very limited expressive capability (e.g., the operator itself is parameter-less) and needs to rely on the interaction arch to encode complex interactions. Moreover, especially when the list is long, using an unweighted sum could deteriorate the signal-to-noise ratio and make the prediction less accurate. By attending to the target user (item), attentive pooling can adaptively adjust how much weight an embedding could get based on not only the relevancy of the current item (user) at hand but also the relevance of other competing entities. This aggregation is especially powerful when the list contains entities of diverse topics (e.g., a user's engagement history could contain items in different categories), for which the multiple distribution modes would be inevitably collapsed into one if sum-pooling is used. Attentive pooling is also more robust and tolerant to noises, outliers or corruptions in the ID list.

\section{Experiments}\label{sec:experiments}

\subsection{On Public Data}
A major goal of this paper is to improve the scaling capability of ranking models due to the curse of quality saturation caused by growing item inventories. To test our findings, data sets need to be both (1) substantially large-scale and (2) based on dynamic inventory as in real-world systems.  Unfortunately, public data cannot meet the requirement: they do not have the desired scale, nor do they have the needed dynamics (matrix completion settings with fixed users \& items). We notice that this is a common issue in the community. Notably, recent works on scaling, including those in NLP and CV are based on dedicated data sets. The matter is even worse in the area of ranking, because published data is not only too small in scale but also lacks many vital characteristics that real-world systems possess, making findings on such toy data sets less reliable when being generalized to real world. However, to improve the reproducibility of our results, we tested our methods on one public data set for demonstration purposes.

\subsubsection{Data}
The MovieLens-20M data set is a popular benchmark in recommendation systems ~\cite{movielens}. It contains  $20$-million ratings from $138,493$ users on $27,278$ movies. In our experiments, we follow a protocol similar to that of \cite{zhou2018deep}: ratings of 4-star or above are treated as positive and the rest as negative; for each user, the most recent $N$ ($N=512$) positively-rated movies are used as item-centric channels of that user; similarly, the $M$ ($M=512$) users who historically rated a movie positively are used as user-centric channels of that movie.
As we mainly compare the difference between ICR and UCR, we do not include other categorical features, such as genre.



\subsubsection{Results}
We tested the DIN \cite{zhou2018deep} architecture (Figure 1(a)) in the three different formulations (i.e, ICR, UCR, hybrid) with `Attention-pooling' as aggregation operator. A 4:1 split is used for training and testing. The evaluation results in terms of AUC (i.e, area under ROC curve) are reported in Table~\ref{tab:movielens}. 

Note that MovieLens is a static data set. It does not have the inventory dynamics that real-world systems have, and hence we will not be able to see parameter explosion on this data set. From Table~\ref{tab:movielens}, our observation is that UCR is at least on par with or slightly better than ICR, while hybrid performs the best possibly because it uses more signals than either of them.
\begin{center}
\begin{table}[!t]
\caption{Evaluation results (AUC) on MovieLens data.}
\centering
\begin{tabular}{|l||*{5}{c|}}\hline
                        &ICR        &UCR        &Hybrid             \\\hline
DIN with Attentive Pooling         &0.712      &0.731      &0.737              \\\hline
\end{tabular}
\label{tab:movielens}
\end{table}
\end{center}

\subsection{On Real-World Production Data}

\subsubsection{Data}
We further experiment on real-world production data.
For offline evaluation, we created a ``lab data set'' by sampling the production log of a real-world short-form video recommendation system. Our data set contains about $24$ million users and their engagement activities in the time range of $60$ days (from late July to early October of $2022$). In total, the data set contains about $28$ billion examples (engagement activities) involving $1$ type of negative and $5$ types of positive engagements.

\subsubsection{Metric}
We use Normalized Cross-Entropy (NCE) as the primary evaluation metric~\cite{He2014PracticalLF}. NCE is defined as the cross-entropy loss of the model prediction $p$ normalized by the entropy of the label $y$.

\begin{align}
NCE(p, y) = \frac{CrossEntropy(p, y)}{Entropy(y)}
\end{align}

NCE is widely used as the gold standard offline metric for engagement probability (e.g, CTR) prediction tasks because of its high consistency with online engagement metrics.



\subsubsection{Parameter Growth}
In both ICR and UCR, the total number of parameters that a model has can be expressed as $const + n\times d$, where the constant part is mostly related to model architectures, while $n$ and $d$ denote the total number of distinctive sparse-ids and the dimensionality of each embedding vector. In our data set, as is common in most ranking systems, the cardinality of the user set tends to be bigger than that of the item set for any given day, $|\mathcal{U}| > |\mathcal{I}_{t+1}| - |\mathcal{I}_{t}|$, where $\mathcal{I}_{t}$ is the accumulative item set on day $t$. However, that comparison is quickly reversed as time goes by because $|\mathcal{I}_{t}|$ grows linearly in $O(t)$. 

\begin{figure}[b]
\centering
\begin{tabular}{ccc}
\includegraphics[width=1.0\linewidth]{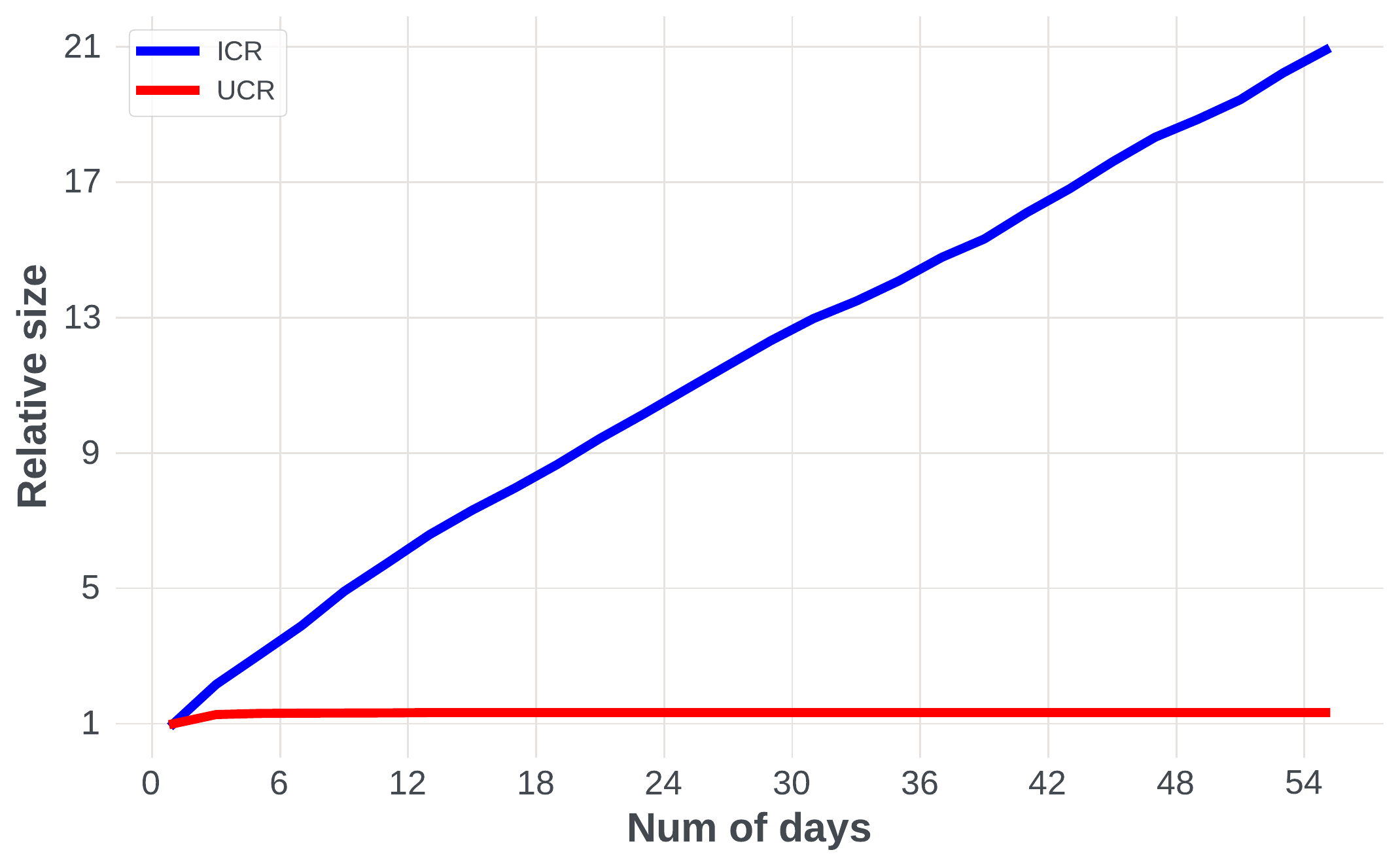} & &
\end{tabular}
\caption{The growths of model size (the total number of parameters) over time for ICR and UCR models.}
\label{FigureModelStats}
\end{figure}

Figure \ref{FigureModelStats} shows the model size growth over time for both ICR and UCR models. We only plotted the curves for the case with sampling and attentive pooling, but the trend is similar for all other variants. While it is true that for the first few days the ICR model has fewer parameters, it constantly adds parameters every day as new item IDs emerge. As a result, the ICR model size grows almost linearly over time. In contrast, the UCR model, although has a bit more parameters initially, the model size stays relatively stable over time.

Considering these two models are trained using the same amount of dyadic interaction data, the drastic contrast of the parameter growth can have profound impacts on model quality. For example, at the end of the 60-day window, the ICR model is 21x larger in size than its UCR counterpart. This means that ICR consumes 21x more memory, or when parameter hashing is used the collision rate is 21x higher; at the same time, on average, each ID embedding receives 21x less training data in ICR as compared to in UCR.
\begin{figure}[t]
\centering
\begin{tabular}{ccc}
\includegraphics[width=1.0\linewidth]{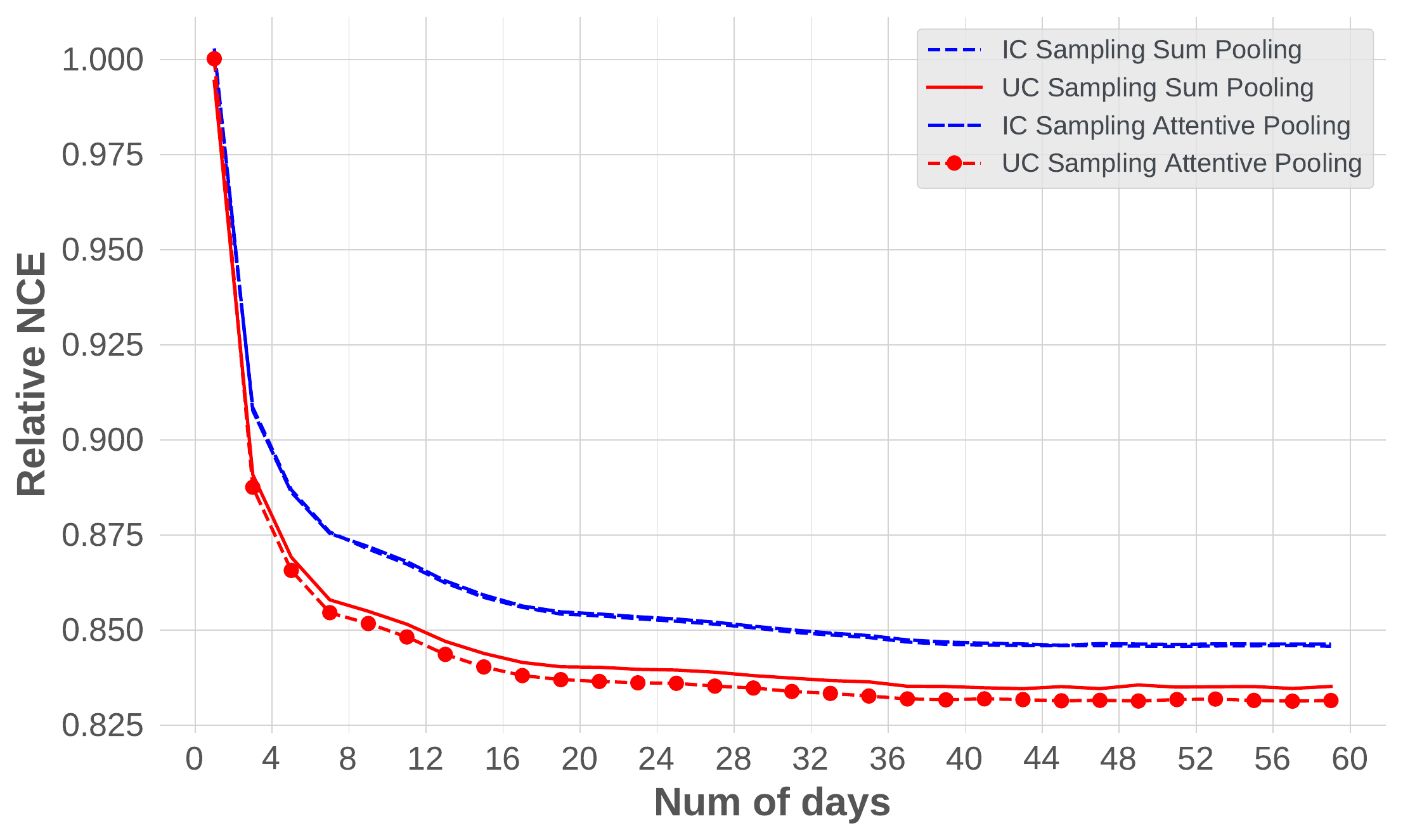} & &
\end{tabular}
\caption{Comparison of ICR and UCR models in offline evaluation. Models are trained recurrently on a daily basis and evaluated on future 10K activities using NCE. (lower is better)}
\label{fig:ic_uc_comparison}
\end{figure}

\subsubsection{ICR vs. UCR}
We compare \texttt{IC-Sampling} and \texttt{UC-Sampling} with the two aggregation operator options. All the models are trained recurrently and evaluated on a daily basis using the first \textasciitilde{}10K examples of the next day. Because we have 6 tasks (and correspondingly 6 engagement history channels) in our data set, each task (and the engagement channel) is evaluated independently. The results are reported in Figure~\ref{fig:ic_uc_comparison}, where only the results on `Task 1' are shown (results on other tasks are very similar); all the NCE numbers are normalized by the NCE of the \texttt{IC-Sampling} sum pooling model on day 1, and relative NCEs are used in the plot. 

We can observe that \texttt{UC-Sampling} demonstrates a clear gain over \texttt{IC-Sampling}, with the gap increasing rapidly from day 1 to day 10, and then slowly converging till the end. The performance matches our hypothesis that UCR accumulates and refines the understanding of each user, which helps with better recommendations as the data scales up. However, we did not notice the gain increase through the end of the experiments. We believe that this is because UCR excels more on active users due to its nature of aggregating user embeddings to profile engaged items, but falls short on less active users. We will come back to address more about this issue in Section~\ref{subsec:segment_analysis}.

\subsubsection{Sum Pooling vs. Attentive Pooling}
We also compare the impact of the two aggregation operators in ICR and UCR. As shown in Figure~\ref{fig:ic_uc_comparison}, attentive pooling consistently performs better than sum pooling in UCR. With more data, the gap is also increasing. After 60 days of training, UCR attentive pooling get 0.44\% gain over the sum pooling alternative. In contrast, the advantage of attentive pooling in ICR is very minimal. 

This also proves our hypothesis in Section \ref{subsubsec:operators}. In ICR, the item ID is not well trained due to the linearly increased ID space . As a result the attention score between history item and target item does not learn useful signals, and attentive pooling falls back to mean (sum) pooling. In UCR, user ID space is stable, and all ID embeddings could be optimized. This finding verifies the potential to solve the quality saturation problem using UCR with more training data.

\subsubsection{Sampling vs. Clustering}
\begin{figure}[t]
\centering
\begin{tabular}{ccc}
\includegraphics[width=1.0\linewidth]{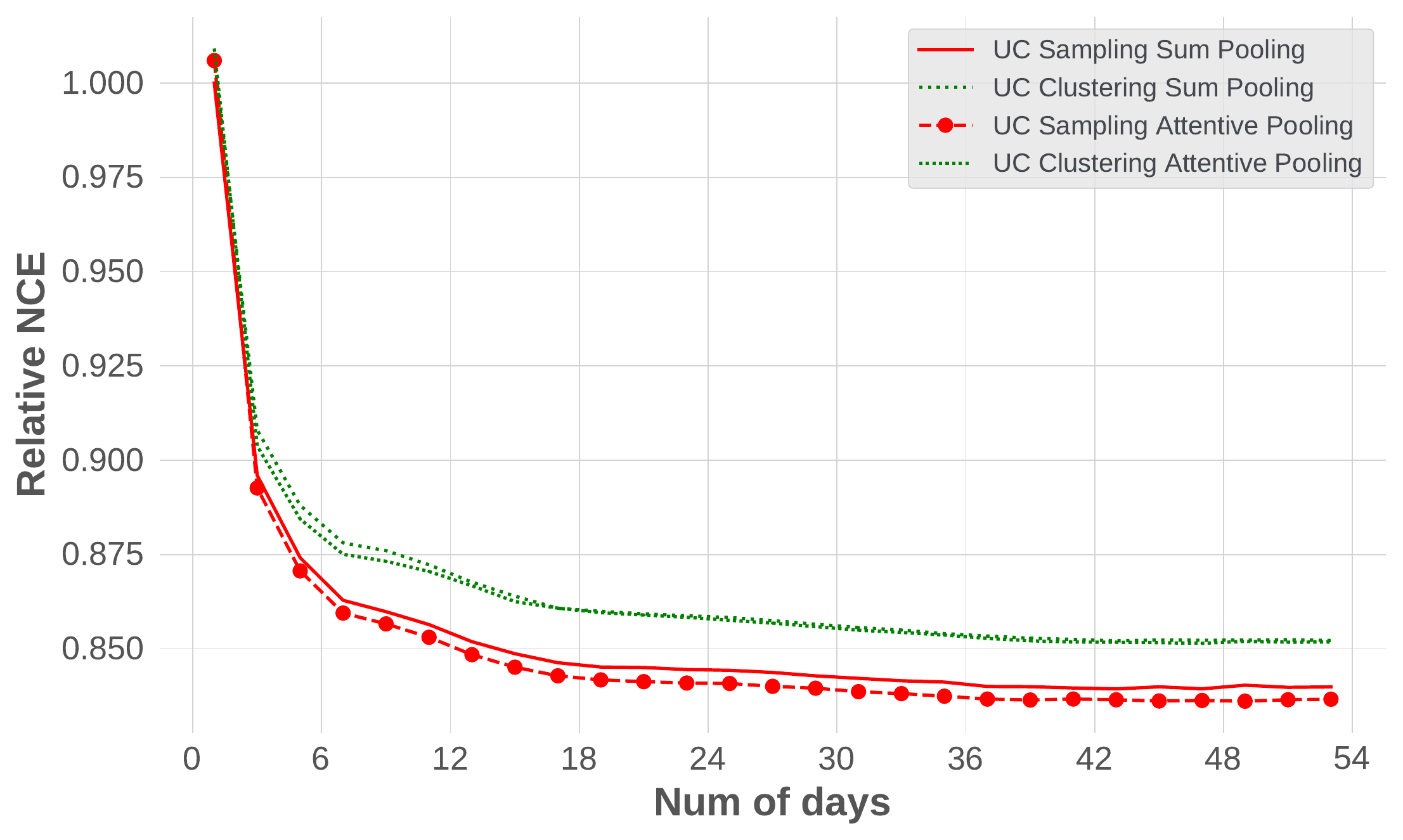} & &
\end{tabular}
\caption{Comparison of the two implementation methods for UCR: sampling vs clustering.}
\label{fig:clustering_comparison}
\end{figure}

In UCR, one of the key aspects to ensure good performance is to construct better and more representative engaged user lists for each item, especially for those extremely popular items that gain millions of user interactions. We implemented two of the approaches presented in Section~\ref{subsubsection:ucr}, namely \texttt{UC-Sampling} and \texttt{UC-Clustering}. Figure~\ref{fig:clustering_comparison} shows the comparison between these two approaches. As can be seen, \texttt{UC-Sampling} seems to dominate \texttt{UC-Clustering} in terms of NCE consistently across the entire time span and all the tasks involved. We want to point out that this may not be definite as the performance highly depends on the choice of implementation, e.g., the incremental Louvain algorithm \cite{blondel2008fast} used in our experiments. If a better algorithm is used, the result can be different. We leave such investigation for future research. 
\begin{figure}[b]
\centering
\begin{tabular}{ccc}
\includegraphics[width=1.0\linewidth]{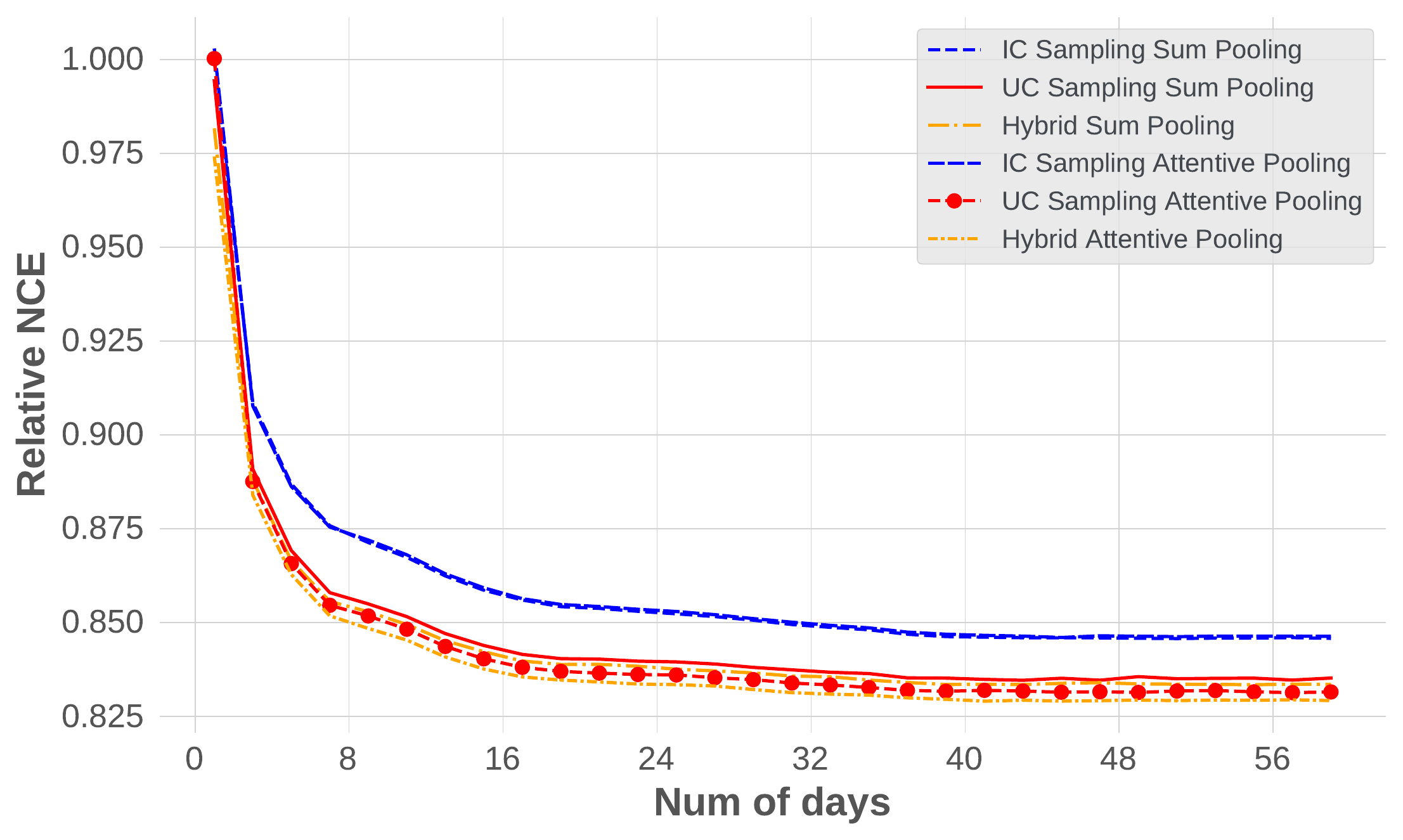} & &
\end{tabular}
\caption{Comparison of the hybrid model with its UCR and ICR counterparts.}
\label{fig:hybrid_comparison}
\end{figure}

\begin{center}
\begin{table*}[!t]
\caption{Multi-task relative NCE percentage (\%) change between ICR (baseline), UCR and Hybrid models implemented with attention pooling. Baseline setting is denoted as ``-".}
\centering
\begin{tabular}{|l||*{12}{c|}}\hline
\multirow{3}{*}[-0.8pt]{Task}  &\multicolumn{3}{c|}{Day 7} &\multicolumn{3}{c|}{Day 14}      &\multicolumn{3}{c|}{Day 30}      &\multicolumn{3}{c|}{Day 60} \\\hline
                    &IC &UC &Hybrid             &IC     &UC     &Hybrid           &IC     &UC     &Hybrid           &IC     &UC     &Hybrid\\ \hline
1                   &- &-2.58 &\textbf{-2.88}   &-1.73  &-4.01  &\textbf{-4.32}   &-3.01  &-4.90  &\textbf{-5.21}   &-3.48  &-5.18  &\textbf{-5.31}   \\\hline
2                   &- &-2.71 &\textbf{-2.94}   &-0.46  &-3.23  &\textbf{-3.42}   &-0.45  &-3.24  &\textbf{-3.44}   &-0.44  &-3.19  &\textbf{-3.32}   \\\hline
3                   &- &+1.84 &\textbf{-2.04}   &-7.52  &-7.60  &\textbf{-10.38}  &-10.96 &-12.64 &\textbf{-14.29}  &-11.98 &\textbf{-13.98} &-12.78  \\\hline
4                   &- &-2.86 &\textbf{-3.08}   &-0.31  &-3.23  &\textbf{-3.41}   &-0.08  &-3.03  &\textbf{-3.23}   &-0.05  &-2.96  &\textbf{-3.08}   \\\hline
5                   &- &-2.88 &\textbf{-3.14}   &-0.66  &-3.61  &\textbf{-3.88}   &-0.79  &-3.77  &\textbf{-4.00}   &-0.81  &-3.73  &\textbf{-3.88}   \\\hline
6                   &- &-3.09 &\textbf{-3.28}   &-0.86  &-3.97  &\textbf{-4.14}   &-1.19  &-4.34  &\textbf{-4.53}   &-1.28  &-4.38  &\textbf{-4.47}   \\\hline
\end{tabular}
\label{tab:all_tasks}
\end{table*} 
\end{center}

\subsubsection{Hybrid Method}\label{subsubsec:hybrid}
We also compare the hybrid method with UCR and ICR. Because the consistently superior performance of sampling over clustering as reported before, we only experimented with the sampling implementation. The results are shown in Figure~\ref{fig:hybrid_comparison}. It seems that the hybrid method has very similar performance as the UCR counterpart, albeit slightly better. This phenomenon is pretty consistent. We observe that the hybrid method achieves the best NCE results across all the tasks. Considering that the hybrid architecture, as shown in Figure~\ref{FigureICR}, includes both an UCR sparse sub-arch and an ICR sparse sub-arch, the results are partly as expected (i.e., it should have the advantages of both UCR and ICR) and partly surprising (i.e., it has the same parameter explosion problem as ICR). 

\subsubsection{Multi-Task Evaluation}\label{subsec:multitask_performance}
In our previous evaluations, we use 1 single task and 1 single engagement history channel. In this section, for both ICR and UCR, we use all the available engagement signal channels (one for each engagement type) and jointly train the model on all of the 6 tasks. This multi-channel and multi-task setting allows the model to capture correlations among different tasks as well as between the signal channel and the task loss corresponding to different engagement types, which cannot be done in the previous setting. The results are reported in Table~\ref{tab:all_tasks}, where the NCE is calculated relative to the NCE of the ICR model at day 7. We observe that overall UCR models show clear gains when compared to ICR counterparts across all the tasks; moreover, the hybrid model consistently performs the best at all of the tasks, although the difference with the UCR models is very marginal. 
\begin{figure}[b]
\centering
\begin{tabular}{ccc}
\includegraphics[height=0.45\linewidth]{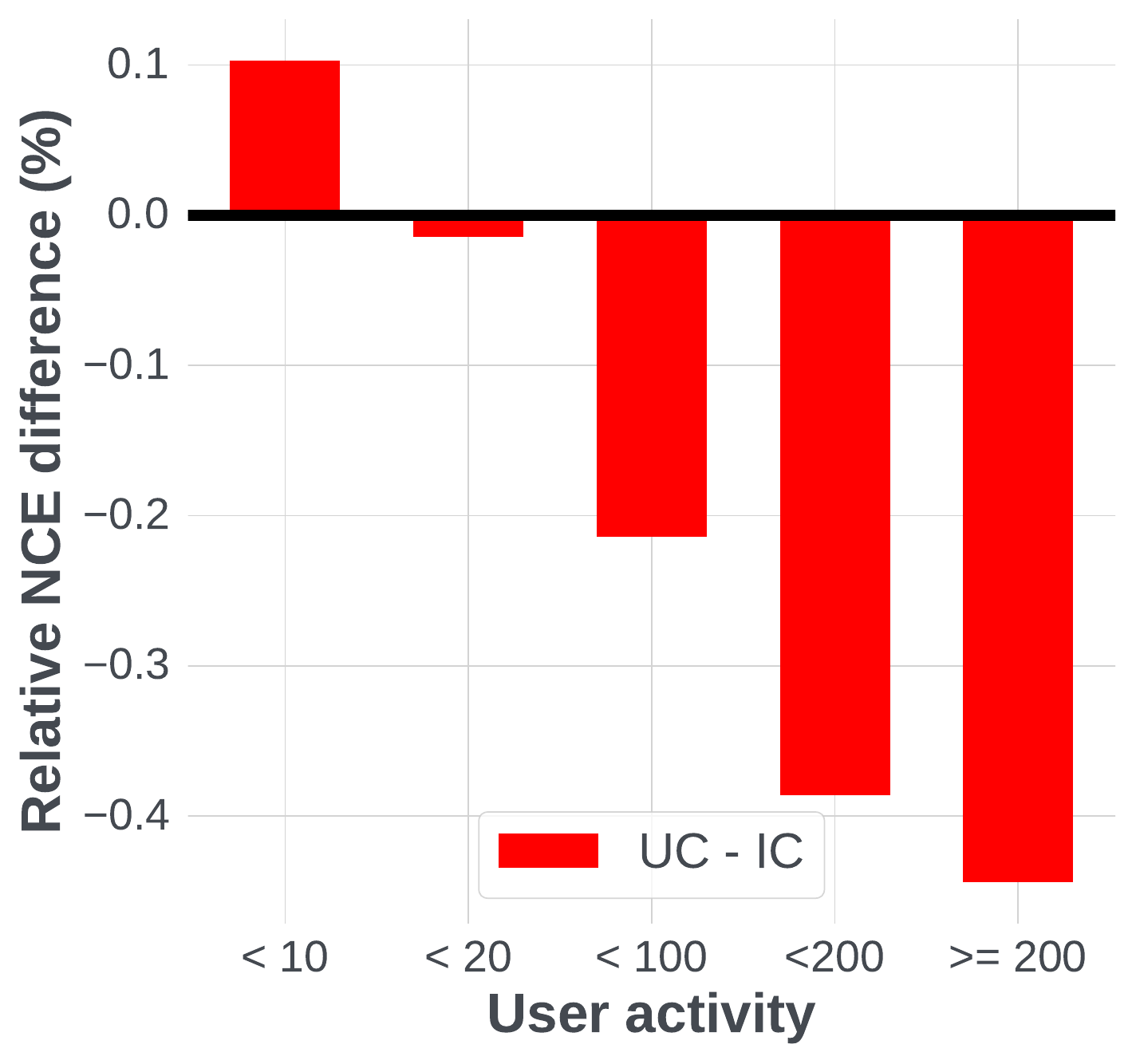} & \hspace{-5pt}
\includegraphics[height=0.45\linewidth]{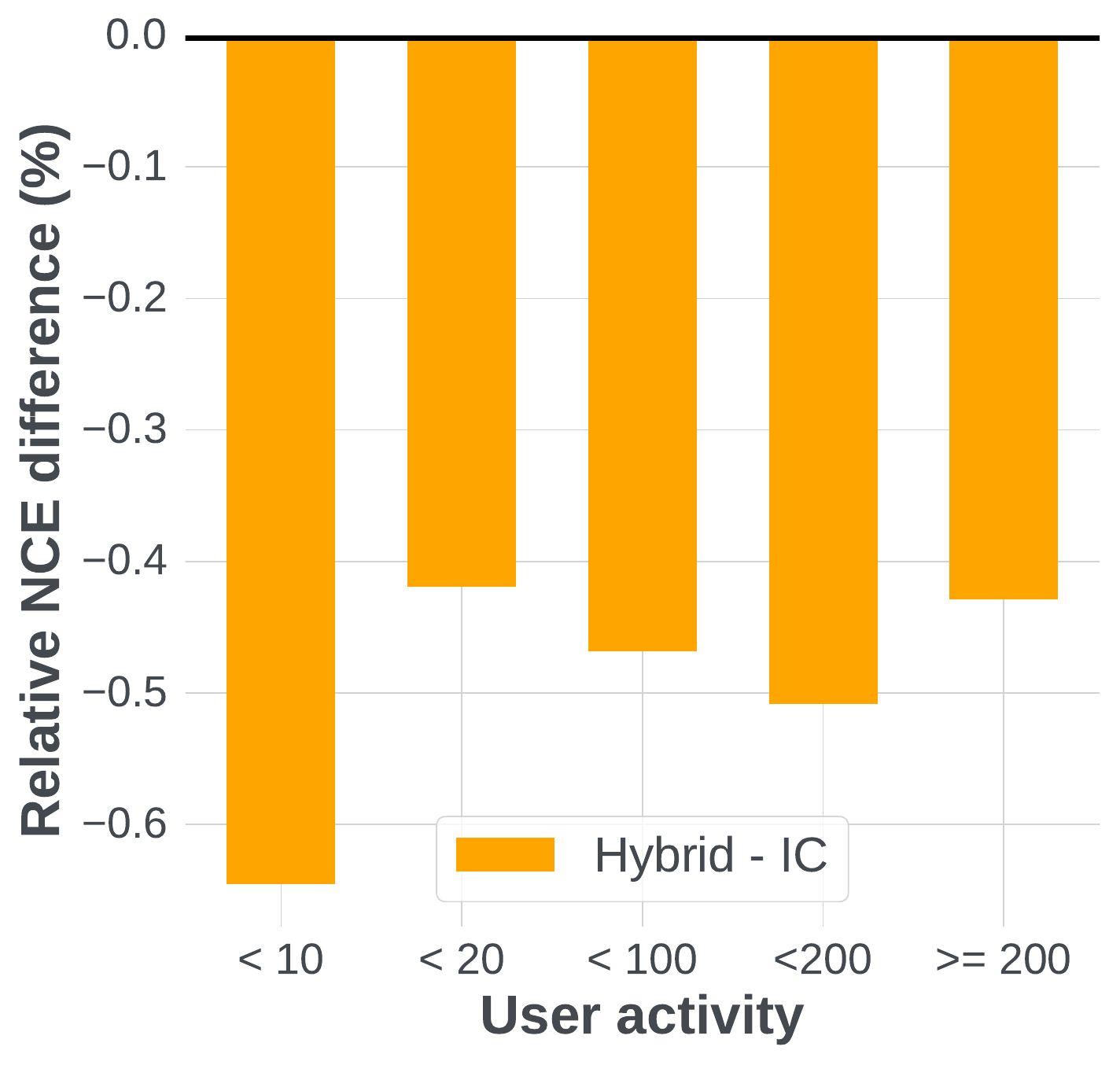} & \vspace{-2pt}\\
~~~~~~~~(a) UCR & ~~~~~~~(b) Hybrid
\end{tabular}
\caption{Distribution of NCE gains over ICR on different user activeness segments (negative means better).}
\label{fig:segment}
\end{figure}

\subsubsection{Segment Analysis}\label{subsec:segment_analysis}
We segment users into five buckets based on their activeness (e.g., number of engagements within a given time window). In Figure~\ref{fig:segment}(a), we show the NCE differences between one UCR model (\texttt{UC-Sampling}) and one ICR model (\texttt{IC-Sampling}) for each user segment. We can see that, although UCR performs better than ICR overall, the gain mostly come from more active users. For less active users (e.g., engagement counts $ < 10$), UCR actually performs worse than the ICR baseline. This explains why the hybrid methods tend to perform the best because it leverages both components to provide the better of the two worlds. As a validation, Figure~\ref{fig:segment}(b) shows the similar analysis of the Hybrid model over ICR, and we can see it provides gains across all the user segments.

\subsubsection{Ablation Study}
To better understand how different configurations impact model performance, we conduct a set of parameter sweep experiments. For this analysis, we set the number of training data to be 30 days for all the runs. In addition, we use \texttt{IC-Sampling} and \texttt{UC-Sampling} with the same single-task setting in our experiments. 



\noindent \textbf{Hash Size.}
Parameter hashing maps user IDs or item IDs to embedding vectors by applying a hash function. Though being space-efficient, it is essential to have a large enough hash space so that a high collision rate between these IDs can be avoided. In this experiment, we further examined how hash size affects model performance by varying it from the default value of 20 million. As hash size affects both IC and UC ranking, we test both \texttt{IC-Sampling} and \texttt{UC-Sampling} as well as using both sum pooling and attentive pooling model architectures. The results are reported in Table~\ref{Tab:hash_size}. Overall, increasing the hash size leads to a better model performance. This trend is more evident for UCR. For example, increasing the hash size from 1M to 30M for UC-Attn results in a $1.71\%$ reduction in relative NCE. One reason why UCR benefits more than ICR is that UCR has much fewer embedding vectors, the reduction in hash collision is more dramatic for UCR when increasing hash size.

%
\begin{center}
\begin{table}[!b]
\caption{Relative NCE percentage (\%) change from different models with varying hash sizes. Baseline setting is denoted as ``-".}
\centering
\begin{tabular}{|l||*{5}{c|}}\hline
        &1M       &5M       &10M      &20M         &30M             \\\hline
IC Sum  &+0.08    &+0.04    &+0.01    &\textbf{-}  &+0.02          \\\hline
IC Attn &+0.12    &+0.05    &+0.05    &+0.07       &+0.06           \\\hline
UC Sum  &-0.04    &-0.73    &-1.07    &-1.43       &\textbf{-1.53} \\\hline
UC Attn &-0.24    &-1.13    &-1.48    &-1.84       &\textbf{-1.95} \\\hline
\end{tabular}
\label{Tab:hash_size}
\end{table}
\end{center}

\noindent \textbf{Embedding Dimensionality.}
We conduct another ablation study on the dimensionality of the embedding vectors. Our default embedding dimension is 192, and we tune it between 96 and 384. Results are illustrated in Table~\ref{tab:feat_dim}. We can see that \texttt{IC-Sampling} is not able to utilize a larger embedding dimension, and its performance is worse when the largest dimensionality is used. On the other hand, \texttt{UC-Sampling} shows consistent improvements when higher dimensional embeddings are used.
\begin{center}
\begin{table}[!t]
\centering
\caption{Relative NCE percentage (\%) change from different models with varying feature dimensions. Baseline setting is denoted as ``-".}
\begin{tabular}{|l||*{3}{c|}}\hline
                &96             &192      &384 \\\hline
IC-Sampling     &-0.05          &-        &+0.01 \\\hline
UC-Sampling     &-1.39          &-1.91    &\textbf{-2.37} \\\hline
\end{tabular}
\label{tab:feat_dim}
\end{table}
\end{center}

\subsection{Online Results}
Based on the encouraging results on the sampled lab data, we took the step forward to productionize the proposed techniques in our recommendation system. \textbf{On the full-scale production data, we observed up to 0.6\% NCE gains} compared to the production ICR model when UCR models were trained with the standard workflow using a few days of training data without any architecture changes. The best version was then tested live in the production system. 

A number of infrastructure optimizations were done to make this happen. For example, we optimize the batching algorithm to put the same user's data in one batch for ICR, so the sum (attention) pooling of the item-centric features only needs to be computed once and then could be shared within the  batch. For UCR, we do the similar operation to batch the same video's data together. With the improvement on data locality, we can lower down the memory consumption, and in turn improve the throughput for both training and serving. Also, by using full-precision for training and lower-precision (e.g., FP16) for inference, we were able to improve the inference performance (both throughput and latency) without significant regression in prediction quality (e.g., NCE) and reduce the number of GPUs required for serving by almost half. The online A/B experiments showed that quite significant wins were achieved across a wide range of topline metrics, in particular, one of the key business metrics, \textbf{video watch time was improved by 3.24\%}.

An important observation during our productionization process is that the offline NCE gain can be further enlarged when we increase the amount of training data. In addition, if we scale up both training data and model complexity, we could potentially obtain an outsized gain in terms of NCE in offline evaluation. This investigation is currently in progress.


\subsection{Open Questions and Discussions}
We are motivated to address the quality saturation problem in ranking. Our expectation is that the UCR formulation should provide somewhat a remedy. However, from our experiment results, this is only partially validated. In particular, we did see UCR models lead to consistently better NCE than their ICR counterparts; we also saw a tendency of improving NCE gain as we increase the training data. Nonetheless, the NCE gap between UCR and ICR is not as big as we expected, and also that gap is being enlarged at a much slower speed, far too slow if we compare it with the model parameter or collision rate growth curves. This is kind of surprising. 

In an attempt to understand the discrepancies, we have a few plausible explanations. 

Firstly, we notice there's a nontrivial discrepancy between the full-scaled production data and our sampled lab data. The scaling characteristics of UCR models are significantly better on production data than what we observed. This is partly related to the sampling algorithm we used to generate this data set, and partly related to the nonlinearity between the complexity that the data manifests and the scale at which the problem is examined.

\begin{figure}[t]
\centering
\begin{tabular}{ccc}
\includegraphics[width=\linewidth]{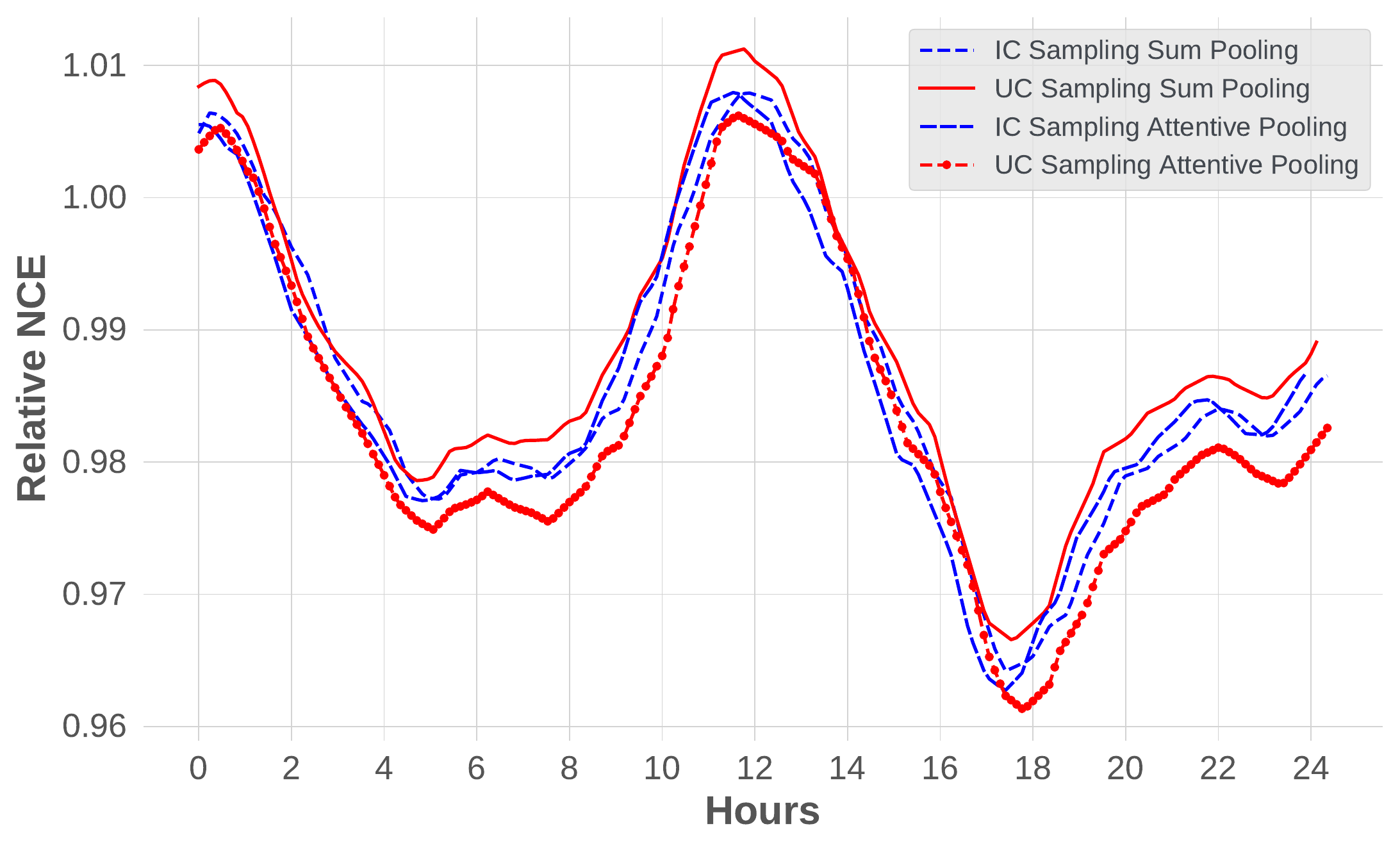} & &
\end{tabular}
\caption{Prediction quality (NCE) of pre-trained models over the next 24 hours indicates there is a strong distribution drift in the data.}
\label{fig:ic_uc_eval}
\end{figure}

Secondly, in the aforementioned areas where scaling has led to tremendous success, including CV and NLP, the concepts we try to model are often static. In other words, there's usually a ground-truth model in hindsight and the goal of training is to approach that ground-truth. However, in ranking it is fundamentally different. There is drastic and frequent distribution drift due to the highly dynamic two-sided ecosystem and the interactive highly counterfactual nature of the engagement process. Because of the distribution drift, there is no ground-truth model (or you could say the optimal model is a moving target instead of static). For example, Figure~\ref{fig:ic_uc_eval} shows how a pre-trained static model performs in the next 24 hours after it was trained. We can see a very significant deterioration of the prediction NCE as the model becomes increasingly outdated. In a situation where the distribution is drifting dynamically, a model that scales well and does not saturate quickly in a static context may not always scale well. To fully combat the obstacles for scaling ranking models, deep understanding of and the ability to control such dynamics are critical.

Last but not the least, our current study is limited, without any changes to the model architecture. We observed, especially for the smaller-scale lab data set, the absolute NCE values are quite small and may be close to their limits for the architecture we used. At the same time, we noticed that ranking model's architectures are significantly simpler than what are commonly used in NLP and CV, which is of course a practical choice given the scales in ranking. We believe that by using significantly more expressive architectures, we will be able to improve the scaling property further.

We leave these investigations for future study.

\section{Summary}
We suspected that the item-centric formulation of ranking models may be contributing to the quality saturation problems. We introduced user-centric ranking as an alternative formulation. We showed that in general, UCR models have a stable model size (i.e., total number of parameters) that will not grow as we increase training data. On a lab data set of sampled production data, we observed that UCR models yield consistently better prediction quality and have slightly better scaling property. We did not believe that this fundamental problem in ranking has been fully solved. We listed a number of open problems from our study and hope they can spark further investigations.

\begin{acks}
We would like to thank the following individuals from Meta for the collaboration and support: Pei Yin, Hui Zhang, Jason Liu, Xianjie Chen, Mingze Gao, Jiyan Yang, Hitesh Kumar, Mert Terzihan, Nathan Berrebbi, Liang Xiong, Jiaqi Zhai, Shilin Ding. Shuang Yang is grateful to Jeff Zheng and Junhua Wang from Newsbreak for many helpful discussions.
\end{acks}


\bibliographystyle{ACM-Reference-Format}
\bibliography{references}

\end{document}